\documentclass[hidelinks,onefignum,onetabnum]{siamart220329}

\usepackage{lipsum}
\usepackage{amsfonts}
\usepackage{bbm}
\usepackage{dsfont}
\usepackage{graphicx}
\usepackage{caption}
\usepackage{subcaption}
\usepackage{epstopdf}
\usepackage{algorithmic}
\ifpdf
  \DeclareGraphicsExtensions{.eps,.pdf,.png,.jpg}
\else
  \DeclareGraphicsExtensions{.eps}
\fi
\newcommand{\Prob}{\mathbbm{P}}

\newcommand{\Pij}{P_{i,j}}

\newcommand{\brac}[1]{\left[#1\right]}

\newcommand{\rvline}{\hspace*{-\arraycolsep}\vline\hspace*{-\arraycolsep}}
\usepackage{amsopn}

\newcommand{\edit}[1]{{\color{black}{#1}}}
\newcommand{\edittwo}[1]{{\color{black}{#1}}}
\newcommand{\editthree}[1]{{\color{black}{#1}}}

\headers{Fast Solver for Diffusive Transport Times on Dynamic Networks}{L. Elam, M. C. Qui\~{n}ones-Fr\'{i}as, Y. Zhang, A. A. Rodal, and T. G. Fai}

\title{Fast Solver for Diffusive Transport Times on Dynamic Intracellular Networks\thanks{
\funding{We acknowledge funding from the Brandeis
NSF MRSEC Bioinspired Soft Materials (NSF-DMR2011846) and NIH R01 NS116375 (AAR and TGF), and NSF grants DMS-1913093 and MCB-2213583 (TGF).}}}

\author{Lachlan Elam\thanks{Department of Mathematics, Brandeis University, Waltham, MA 
  .}
\and M\'onica C. Qui\~{n}ones-Fr\'{i}as\thanks{Department of Biology, Brandeis University, Waltham, MA.}
\and Ying Zhang\footnotemark[2]
\and Avital~A.~Rodal\footnotemark[3]~\and~Thomas G. Fai{\footnotemark[2] \thanks{\email{tfai@brandeis.edu}}}}

\ifpdf
\hypersetup{
  pdftitle={Fast Solver for Diffusive Transport Times on Dynamic Intracellular Networks},
  pdfauthor={L. Elam, M. Qui\~{n}ones Fr\'{i}as, Y. Zhang, A. A. Rodal, and T. G. Fai}
}
\fi

\begin{document}

\maketitle
\begin{abstract}
     The transport of particles in cells is influenced by the properties of intracellular networks they traverse while searching for localized target regions or reaction partners. Moreover, given the rapid turnover in many intracellular structures, it is crucial to understand how temporal changes in the network structure affect diffusive transport. In this work, we use network theory to characterize complex intracellular biological environments across scales. We develop an efficient computational method to compute the mean first passage times for simulating a particle diffusing along two-dimensional planar networks extracted from fluorescence microscopy imaging. We first benchmark this methodology in the context of synthetic networks, and subsequently apply it to live-cell data from endoplasmic reticulum tubular networks.
\end{abstract}



\section{Introduction}
    Previous experimental and theoretical studies reveal how the diffusive exploration of objects relates to the morphological properties of intracellular environments \cite{ZaniEdelman2010,Masudaetal2017,Benichouetal2011,Kosloveretal2011}. One particularly important category of intracellular diffusive transport emerges on network-like structures used to describe the morphologies of subcellular organelles \cite{Vianaetal2020}. Lipids and proteins are synthesized in the ER, and their rate and direction of transport from their site of synthesis to other points in the ER network, where they can be delivered to other organelles, remain poorly understood. Objects traversing along the edges of a network exhibit a broad range of transport behaviors dependent on the properties of the network. For example, previous work has demonstrated the importance of the edge-length distribution in overall search times on networks through analyzing the mean first passage time (MFPT) on planar networks \cite{Brownetal2020}. Motivated by the rapid turnover observed in subcellular structures, in this work we develop an efficient computational method to compute how changes in network morphology affect the diffusive MFPT.

Changes in the structure of confined networks have been shown to have a significant impact on transport in a broad variety of biological problems, including cytoskeletal intracellular transport \cite{Andoetal2015} and cargo trafficking along the tubules of the endoplasmic reticulum (ER) \cite{Niuetal2019}. In this work, we use a coarse-grained model of random walks on networks to simulate the diffusion on networks. By developing an efficient approach based on the Sherman-Morrison formula, we are able to calculate how changes in the network affect transport. In particular, we focus on ER networks obtained from experiments and apply the method to show how the properties of the network influence the MFPT of particles diffusing along ER tubules. We find that changes in the ER network structure within dense and highly-connected regions (e.g.~in a neighborhood around the nucleus) tend to speed up the diffusive search of a particle. On the other hand, the loss of tubules that bridge together different components of the network tends to slow down the search process.

An open-source repository with implementations of the methods described here is available at \url{https://github.com/Lanlach1/SIAM-Diffusive-Transport-Time-Solver}.

\section{Mathematical Model}
\subsection{Transition probabilities along edges of a static network}
\label{sect:TransitProb}
The networks that we consider are a system of nodes connected by edges with each node having $N_i$ ($N_i \geq 1$) connections and particles searching along a network via transiting from nodes. Assuming particles undergo an unbiased random walk along the edges of the network, the transition probabilities satisfy \cite{Koslover2012,Brownetal2020,PetterutiThesis}:
\begin{equation}
    \Pij = \frac{1}{l_j}\frac{1}{\sum_{j = 1}^{N_i}\frac{1}{l_j}},
    \label{eq:Pi_4}
\end{equation}
where $\Pij$, $j = 1, 2, \dots, N_i$ denote the transition probabilities from $X_i$ to its $j^\text{th}$ connected node, $X_j$, and $l_j$ is the length of the edge between $X_i$ and $X_j$.

\editthree{Similarly, we define $Q_i$ as the average time spent at node $i$ prior to stepping to one of its neighbors, which from \cite{Brownetal2020} satisfies
\begin{equation}
    Q_i = \frac{1}{2D}\frac{\sum_{j = 1}^{N_i} l_j}{\sum_{j = 1}^{N_i} l_j^{-1}},
\end{equation}
where $D$ is the diffusion coefficient. In this work, we take $D=1$ \textmu m\textsuperscript{2}/s throughout, which gives an order of magnitude estimate for typical proteins diffusing in cells \cite{milo2015cell}.}

\editthree{We define the stationary distribution $\mathbf{z}\in\mathbb{R}^{n}$ from the transition matrix $P$ such that $\mathbf{z}^T P = \mathbf{z}^T$.  As defined, $P$ is a row stochastic matrix, therefore we may compute $\mathbf{z}$ by solving for the left eigenvector of $P$ associated with eigenvalue 1. Each node $i$ will have steady state probability $z_i$.}

\subsection{Mean first passage time}
\label{sect:MPFT}
We quantify the number of nodes visited on average during diffusion between any two nodes across the network using the mean first passage time (MFPT), which can be computed from the transition matrix $P$ with entries given by \cref{eq:Pi_4}. It follows from the Markov property of the random walk that the transition probability of reaching node $X_i$ from $X_j$ in two steps is given by
\begin{align*}
        \Pij^2 = \sum_k P_{ik}P_{kj}= \Prob\brac{\text{Transition to }X_i \vert \text{Current Node} = X_j\, \text{and}\, 2 \,\text{steps}}.
\end{align*}
We may further generalize this property to obtain the transition probability of reaching node $i$ from node $j$ in 3 steps:
\begin{align*}
        &\Pij^3 = \sum_h P_{ih} \sum_k P_{hk}P_{kj}= \Prob\brac{\mbox{Transition to} \, X_i \, \vert \, \mbox{Current Node} = X_j \,\text{and} \, 3\, \text{steps}}.
\end{align*} 
It follows by induction that
\begin{equation}
    \begin{aligned}
        \Pij^N = \Prob[\text{Transition to} \, X_j \, \vert \, &\mbox{Current Node} = X_i \,\text{and}\, N\, \text{steps}].
    \end{aligned}
    \label{eq:PijN}
\end{equation}

\Cref{eq:PijN} gives us the general formula for the probability of a particle to travel diffusively from node $i$ to node $j$ in $N$ steps. 

To compute the expected value of the average number of steps taken during diffusion between nodes $i$ and $j$ using \cref{eq:PijN}, we must specify a new transition matrix in which there is zero probability of leaving the target node once the particle has arrived. We denote by $\edit{P_r} \in \mathds{R}^{(n-1)\times(n-1)}$ the new transition matrix obtained by removing the $\edit{r}^{th}$ row and column of the original matrix $P$, where $\edit{r}$ is our target node. $\edit{P_r}$ is constructed as follows:
\begin{equation}
\edit{P_r} = 
\begin{pmatrix}
  \begin{matrix}
  P_{1:\edit{r}-1, 1:\edit{r}-1}
  \end{matrix}
  & \rvline & P_{\edit{r}+1:n, 1:\edit{r}-1} \\
\hline
  P_{1:\edit{r}-1, \edit{r}+1:n} & \rvline &
  \begin{matrix}
  P_{\edit{r}+1:n, \edit{r}+1:n}
  \end{matrix}
\end{pmatrix}.
\label{eq:P_t}
\end{equation}
From here we may generate an expected value of the amount of time spent on node $j$ given initial condition $i$ and target node $\edit{r}$:
\begin{equation}
    \begin{aligned}
      \brac{ \; \sum^\infty_{k = 0} P^k_\edit{r}}_{i,j} &=  [I + \edit{P_r} + \edit{P_r}^2 + \edit{P_r}^3 ...]_{ij} =  [(I - \edit{P_r})^{-1}]_{ij}, \\
      &= \mathbb{E}[\mbox{\# of visits to $j$ } | \mbox{ Initial Node} = i] =: \left(\edit{N_r}\right)_{i,j},
    \end{aligned}
    \label{eq:Nt}
\end{equation}
where the matrix $\edit{N_r}$ denotes the fundamental matrix for target condition $\edit{r}$. The $(i,j)^\text{th}$ element of $\edit{N_r}$ represents the average number of steps a particle will take to node $j$ while diffusing from $i$ to $\edit{r}$. Note that the formula for $\edit{N_r}$ requires a matrix inversion, for which we will consider efficient  computational strategies later on.

The MFPT is then obtained by summing over row $i$ of $\edit{N_r}$ to find the \editthree{average amount of time} spent diffusing over the entire network given initial condition $i$ and target condition $\edit{r}$:
\begin{equation}
\editthree{    \edit{M_{ir}}:= (\edit{N_r}\cdot \mathbf{Q})_i = \sum_j \mathbb{E}[\text{\# of visits to $j$ } \vert \text{ Initial Node} = i] \cdot\mathbb{E}[\text{time at $j$}]},
    \label{eq:Mit}
\end{equation}
which yields the MFPT matrix $M \in \mathbb{R}^{n\times n}$. The $(i,r)^\text{th}$ entry of $M$ gives the MFPT to go from node $i$ to node $\edit{r}$.

\edit{
\subsection{Mean transit time}
}
The MFPT matrix may be used to assess the average time required to reach a given node. To compute the \edit{mean transit time, or global mean first passage time (GMFPT) to a node, we average the MFPT to that node with equal weight over all initial conditions, following \cite{zhang2009random,zhang2009influences,tejedor2009global,Brownetal2020}}. In particular, to compute the mean transit time to node $r$, we set it as the target condition as described above. 
Letting $\widehat{\mbox{M}}_i$ denote the $i^{th}$ row of the MFPT matrix, the GMFPT to node $r$ is given by:
\begin{equation}
    C_r := \frac{1}{n}\sum_i\mathbb{E}[\mbox{travel time to $r$ } | \mbox{ Initial Node} = i] = \frac{1}{n}\cdot\widehat{\mbox{M}}_r^T\cdot \mathds{1}.
    \label{eq:connectivity}
\end{equation}

\edit{To provide intuition for this formula, note that the GMFPT is simply the average of the MFPT of all possible initial conditions to arrive at node $r$. The $C_r$ values at different nodes give a measure of their relative integration within the network, i.e.~smaller values of $C_r$ correspond to nodes that are easily reached from the rest of the network, whereas larger values of $C_r$ correspond to nodes that take longer times to reach.}

\subsection{Edge Importance}
In addition to the \edit{mean transit time}, we may use the MFPT to evaluate the importance of a given connection between two nodes. To obtain a quantity that characterizes the overall network, we average $C_r$ over all nodes $r$. This quantity was introduced previously in \cite{Koslover2012}, where it is referred to as the Target Averaged Global Mean First Passage Time (TAGMFPT). In the notation of this article, the TAGMFPT is defined as 
\begin{equation}
    \mbox{TAGMFPT}:=\frac{1}{n}\sum_r C_r = \frac{1}{n^2}\sum_r \edit{\widehat{\mbox{M}}_r^T}\cdot \mathds{1}.
    \label{eq:GMFPT}
\end{equation}
We may recompute the TAGMFPT for the same network but after removing the connection between nodes $i$ and $j$. We shall call this new value $\mbox{TAGMFPT}^\star_{i,j}$, as it is derived using $\edit{N^\star_r}$ described in \cref{eq:N_star}, where $\edit{N^\star_r}$ is the given fundamental matrix for target condition $\edit{r}$ for our network with one connection removed. We can look at the difference between these two values to assess the importance of the connection should it be removed from the network:
\begin{equation}
    E_{i,j} := \mbox{TAGMFPT}^\star_{i,j} - \mbox{TAGMFPT},
    \label{eq:E}
\end{equation}
where $E_{i,j}$ is the importance of the connection between nodes $i$ and $j$. For intuition, if $E_{i,j} > 0$ we can assume that removing the $i,j^{th}$ edge makes the average amount of steps taken for any path increase.

\subsection{Synthetic Networks}
\label{sect:artificialNetwork}
\begin{figure}[!htbp]
    \centering
    \includegraphics[width=\textwidth]{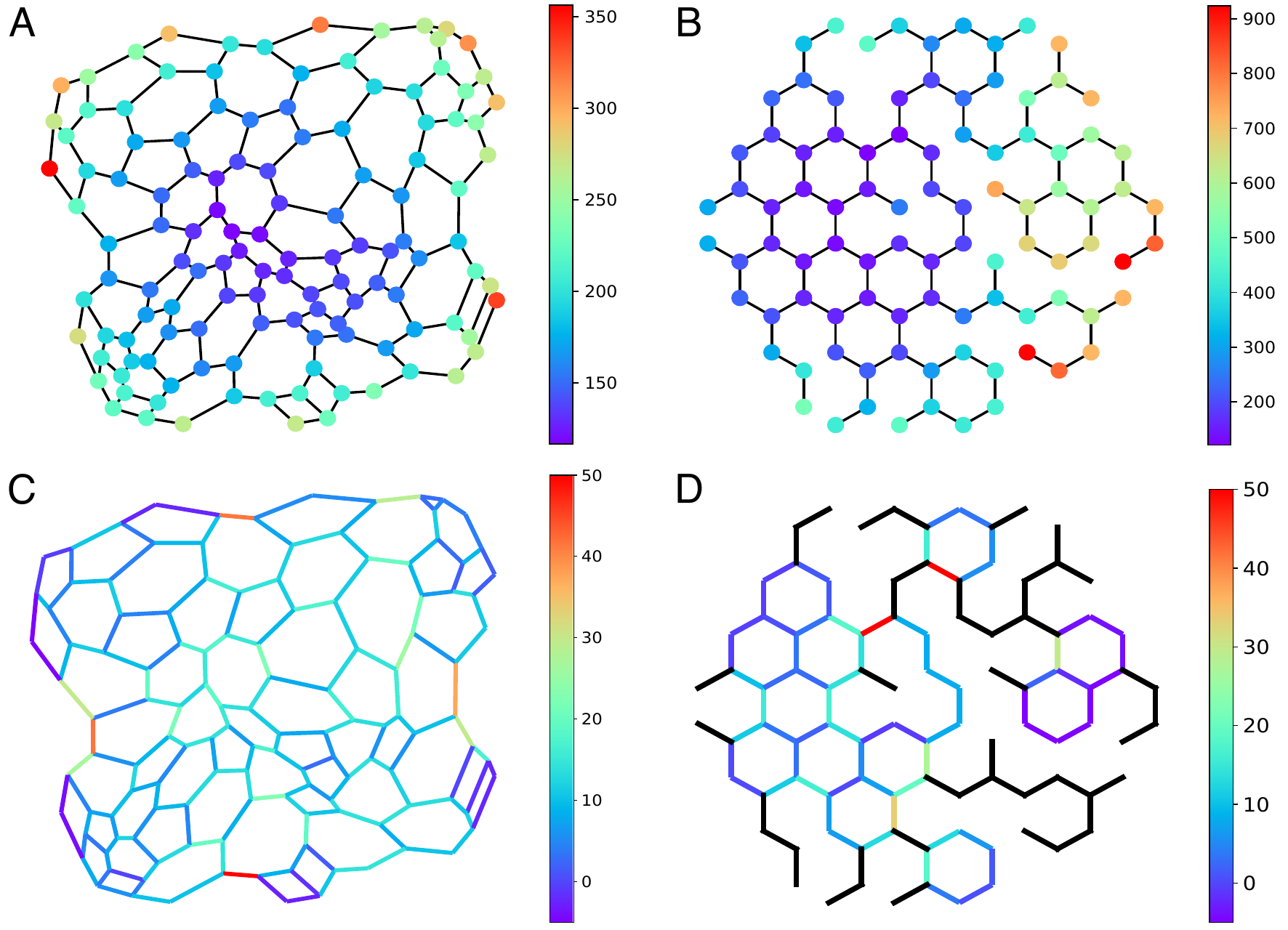}
    \caption{\edit{Global mean first massage time (GMFPT)} and edge importance of synthetic networks. (A and B) The \edit{GMFPT} for the Voronoi network (A) and the augmented honeycomb network (B). (C and D) Edge importance for the Voronoi network (C) and the augmented honeycomb network (D). \edit{Critical edges in panel (C) and (D) are labeled in black. The length scale for each network is chosen so that the mean edge length is 1 \textmu m.} \editthree{Here and throughout the manuscript, times are given in units of seconds.}}
  \label{fig:FIG01}
\end{figure}

To benchmark the method, we first applied it to various synthetic networks. We created simple networks using Delaunay triangulation of a random distribution of points within a rectangular domain, and then took the Voronoi network (the dual graph). These networks bear some similarities to biological ER networks in that the majority of junctions within the network have degree three. Moreover, the large porous areas created by rings of junctions are reminiscent of the ER morphology. These properties make them a useful benchmark for our purposes. In \cref{fig:FIG01} we show an example of these synthetic networks.

We begin by testing our method on networks with a fixed number of connections per node to quantify how the degree of a given junction affects MFPTs across the network. The Voronoi network as shown in \cref{fig:FIG01}A is very close to having all degree-three nodes, except for nodes on the boundary that may have fewer edges. 

As a benchmark comparison similar to that reported in \cite{Koslover2012,Brownetal2020}, we generated an augmented, non-symmetric honeycomb structured network through randomly removing connections from its corresponding symmetric form (see \cref{fig:FIG01}B). These honeycomb networks have a propensity to form long and isolated branches, which are significant when testing nodes for \edit{GMFPT}.
The honeycomb networks are created by first generating the centroid of each hexagon within the network. The middle row of hexagons is always made odd to preserve symmetry of the network.

To analyze the list of synthetic networks described above, we make use of the MPFT described in \cref{sect:MPFT}. For the Voronoi network (\cref{fig:FIG01}A), we quantified its \edit{GMFPT} using \cref{eq:connectivity} and found that the \edit{GMFPT} of a given node is most dependent on centrality (distance from the center of network). This is due to the fact that these networks are mainly composed of degree three nodes and do not form isolated branches extending from the main network.

The augmented honeycomb network (\cref{fig:FIG01}B) has a feature that distinguishes it from the other synthetic networks. For this particular type of network, all features are the same as the original honeycomb except edges were removed with a probability of 0.2. This means that a node within the interior of the network has a $0.2$ probability to have two connections, and $0.2^2 = 0.04$ probability to have only one connection. This creates large isolated branches within the network, which heavily influence the \edit{GMFPT} per node (see \cref{fig:FIG01}B), while \edit{maintaining overall network properties such as the TAGMFPT}.

To further investigate the importance of an edge, we repeat the centrality analysis. For the Voronoi network we find that the importance of an edge is mainly influenced by two traits: 1) how central the connection is, and 2) if removing that connection will increase the minimum path between any two nodes greatly (see \cref{fig:FIG01}C).

The augmented honeycomb network is unique from the synthetic Voronoi network. This is because there exist connections within the network such that if they were to be removed, two disconnected networks would be created. We shall call these connections critical edges. We can check for these critical edges by solving for the condition \cref{eq:eq5} as described in \cref{sect:ShermanMorrison}. In \cref{fig:FIG01}D we label these edges in black.

\subsection{The Sherman Morrison formula}
\label{sect:ShermanMorrison}
To study the effect of temporal changes in the network, such as removal of an edge, in practice one needs an efficient method to compute the MFPT matrix \cref{eq:Mit}. The Sherman-Morrison (SM) formula is a well-known formula for the inverse of a matrix with a rank-1 update. Using this formula we may efficiently invert perturbations of the MFPT matrix e.g.~for swapped target conditions. In particular, given vectors $u,v \in \mathds{R}^n$ and an invertible matrix $A \in \mathds{R}^{n\times n}$ with $(A + uv^T)^{-1}$ invertible, the SM formula yields the inverse of $A + uv^T$ as
    \begin{equation}
        (A + uv^T)^{-1} = A^{-1} - \frac{A^{-1}uv^TA^{-1}}{1+v^TA^{-1}u}
        \label{eq:SM1}
    \end{equation}
Note the assumption that $(A + uv^T)^{-1}$ is invertible, or equivalently ${1+v^TA^{-1}u} \neq 0$ upon taking the determinant. We will discuss the importance of this criterion later on. 

We denote by $\edit{N_r}$ the fundamental matrix of our original network and $\edit{N^\star_r}$ the fundamental matrix of a network with one edge removed. Using \cref{eq:Nt} we can make a series of rank 1 updates to $\edit{P_r}$ to create $\edit{P^\star_r}$, and then using the vectors of the rank 1 updates, we can apply the SM formula to compute the same rank one updates on the inversion $\edit{N_r}$ to create $\edit{N^{\star}_r} = (I - \edit{P^\star_r})^{-1}$.
For a simple change like an edge removal, our transition matrices for each network, $\edit{P_r}$ and $\edit{P^\star_r}$ are very similar. If an edge was removed between nodes $i$ and $j$, then the difference between $\edit{P_r}$ and $\edit{P^\star_t}$ would only be in the $i^{th}$ and $j^{th}$ rows and columns. We let $e_k \in \mathds{R}^{n-1}$ be the $k^{th}$ standard basis vector and $p_i, p^\star_i \in \mathds{R}^{n-1}$ be the $i^{th}$ column of $\edit{P_r}$ and $\edit{P^\star_r}$ respectively. Similarly, we denote by $b_i, b^\star_i \in \mathds{R}^{n-1}$ the $i^{th}$ rows of $\edit{P_r}$ and $\edit{P^\star_r}$ respectively. Given that the only difference between $\edit{P_r}$ and $\edit{P^{\star}_r}$ is to the rows and columns associated with the nodes connected to the edge we wish to remove, we can easily equate the two. This is accomplished by swapping out the necessary columns and rows: 
\begin{equation}
   \edit{P^\star_r} = \edit{P_r} + (p^\star_i - p_i)e^T_i +(p^\star_j - p_j)e^T_j + e_i(b^\star_i - b_i)^T +e_j(b^\star_j - b_j)^T
    \label{eq:P_star}
\end{equation} 
We subsequently introduce two vectors $\rho_i = p^\star_i - p_i$ and $\beta_i = b^\star_i - b_i$ to rewrite \eqref{eq:P_star} in a simpler form:
\begin{equation}
  \edit{P^\star_t} = \edit{P_t} + \rho_ie^T_i +\rho_je^T_j + e_i\beta_i^T +e_j\beta_j^T.  \label{eq:P_star_small}
\end{equation}
We note that \eqref{eq:P_star_small} is now a series of rank 1 updates on $\edit{P_r}$ and the fundamental matrix $\edit{N^\star_r}$ may be calculated by
\begin{equation}
    \edit{N^\star_r} = (I - \edit{P_r} - \rho_ie^T_i -\rho_je^T_j - e_i\beta_i^T -e_j\beta_j^T)^{-1}.
    \label{eq:N_star}
\end{equation}
We can then apply the SM formula to extract $\edit{N^\star_r}$ from $\edit{N_r}$ in a computationally efficient manner, using the vectors provided. From here we may easily compute $M^\star$, which is our MFPT matrix for the same network except with our desired edge removal.

When dealing with inverting a matrix we must also check that the matrix is indeed invertible. We must check that the determinant of our new matrix satisfies
    \begin{equation}
        \det(A + uv^T) = (1+v^TA^{-1}u) \det(A).
        \label{eq:Matrix_det_lemma}
    \end{equation}

Given that $\edit{N^\star_r}$ is a series of rank 1 updates on $\edit{N_r}$ as described in \cref{eq:N_star}, we may compute the determinant:
\begin{equation}
    \det(I - \edit{P^\star_r}) = (1 + e^T_i\edit{N_r}^1\rho_i)(1 + e^T_j\edit{N_r}^2\rho_j)(1 + \beta_i^T\edit{N_r}^3e_i)(1 + \beta_j^T\edit{N_r}^\star e_j)\det(I - \edit{P_r}).
    \label{eq:det_n_star}
\end{equation}
Note that $\edit{N_r}^1, \edit{N_r}^2, \edit{N_r}^3, \edit{N_r}^\star$, define the fundamental matrix after each rank one update to $I-\edit{P_r}$. We assume that $\edit{N_r}$ is invertible, which is equivalent to $\det(I - \edit{P_r}) \neq 0$, meaning that the original network is irreducible. This simplifies the condition of checking if the new $\edit{N^{\star}_r}$ is invertible as we only need to show that:
\begin{equation}
    (1 + e^T_i\edit{N_r}^1\rho_i)(1 + e^T_j\edit{N_r}^2\rho_j)(1 + \beta_i^T\edit{N_r}^3e_i)(1 + \beta_j^T\edit{N_r}^\star e_j) \neq 0
    \label{eq:eq5}
\end{equation}
Given that we only need to check the rank one updates individually for invertibility, we may simplify \eqref{eq:eq5} by replacing $\edit{N_r}^1, \edit{N_r}^2, \edit{N_r}^3, \edit{N_r}^\star$ with $\edit{N_r}$.
\begin{equation}
    (1 + e^T_i\edit{N_r}\rho_i)(1 + e^T_j\edit{N_r}\rho_j)(1 + \beta_i^T\edit{N_r}e_i)(1 + \beta_j^T\edit{N_r}e_j) \neq 0
    \label{eq:eq6}
\end{equation}
We note that \cref{eq:eq6} does not hold in general. If the criterion \cref{eq:eq6} is not satisfied, we deduce that the connection removed to create $\edit{N^\star_r}$ results in a reducible network with two disconnected sub-networks. However, there is one exception: if a connection is removed that only separates one node from the rest of the network, then the deduction is false because in the case of a disconnected network \cref{eq:eq6} does not hold. However, we may account for this special case by first testing if removing a given edge results in a degree zero node. This is computationally inexpensive, and the combination of checking if \cref{eq:eq6} is true and that no nodes are degree zero is sufficient to conclude that the new network is irreducible. 

The SM formula may also be used to efficiently compute all possible target conditions of the fundamental matrix. This is explained in detail in \cref{app:sm}.
\begin{figure}[!htbp]
    \centering
    \includegraphics[width=\textwidth]{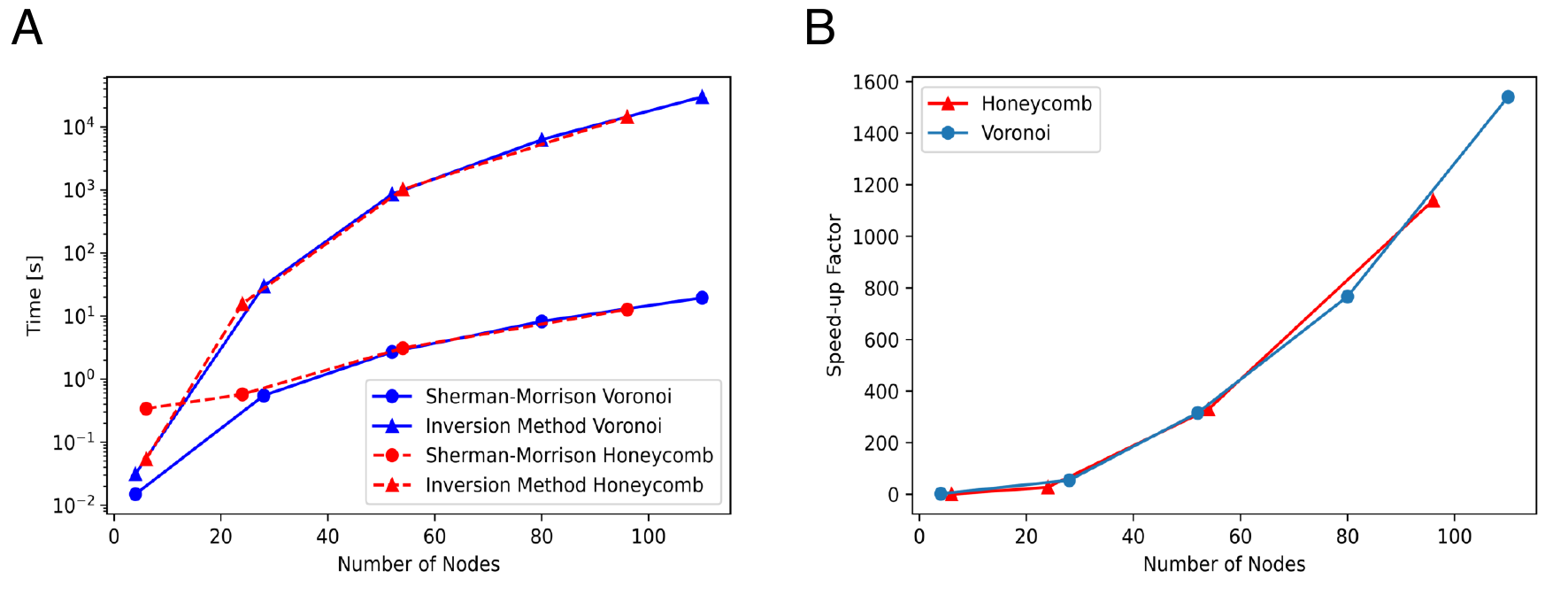}
    \caption{Computation time and efficiency comparison of the original method and the optimized method using the SM formula. (A) Computation time of both methods for the Voronoi network (blue) and the augmented honeycomb network (red) with varying complexity. The solid lines with triangle markers correspond to the original method without using the SM formula. The solid lines with circle markers correspond to the optimized method using the SM formula. (B) The corresponding speed-up factor for the Voronoi network (blue-circle) and the augmented honeycomb network (red-triangle) calculated using data shown in (A).}
  \label{fig:FIG02}
\end{figure}

Using these two applications of the SM formula, we now only need to directly invert a single matrix, and from this single matrix inverse we can compute any target condition or simple network change without having to directly invert a new matrix. To demonstrate the efficiency of this method, we compared the cost of computing the MFPT matrix using both the original and optimized methods on networks of increasing size. In \cref{fig:FIG02}A, we show that the computation time increases significantly using the original method as the number of nodes is increased for both the Voronoi and the augmented honeycomb networks; whereas the computation time remains small using the optimized method. To further quantify the difference between the two methods, we calculated the speed-up factor for each network complexity. Here the speed-up factor is defined to be the ratio of the slower to the faster computation time. For both network types, we see a significant increase in the computation speed using the optimized method (\cref{fig:FIG02}B) as the network becomes more complex.

\section{Application to ER networks}
\label{sect:ERnetwork}
To explore the importance of network properties in dynamic biological organelles, we applied MFPT analysis to a series of ER networks. These networks expand over an entire cell and consist of hundreds of nodes and thousands of connections. Instead of attempting to accurately model the network remodeling dynamics, here we extract the network dynamics directly from data and compare the observed dynamics to an extremely simple version of remodeling in which only single edge deletion is allowed. Recall that the difference in diffusive transport time upon edge deletion is precisely the quantity measured by the edge importance introduced previously.
\begin{figure}[!htbp]
    \centering
    \includegraphics[width=\textwidth]{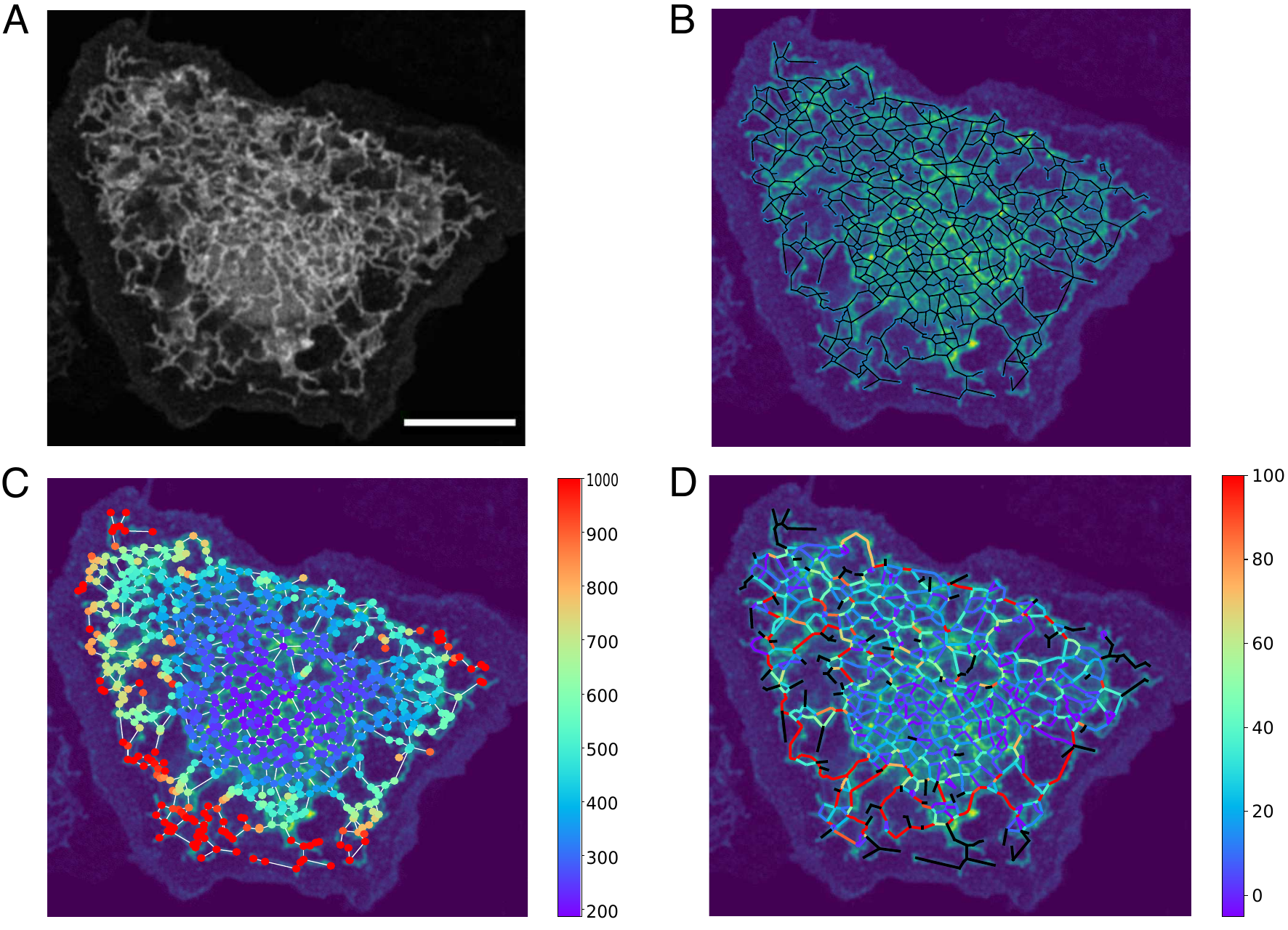}
    \caption{Diffusive transport analysis of an S2 cell. (A) The original cell image and the corresponding (B) extracted network, (C) \edit{GMFPT}, and (D) Edge importance. \edit{A 5 micron scale bar is plotted at the right bottom corner in panel (A). The critical edges in panel (D) are shown in black (see also \cref{app:S3}).}}
  \label{fig:FIG03}
\end{figure}

\begin{figure}[!htbp]
    \centering
    \includegraphics[width=\textwidth]{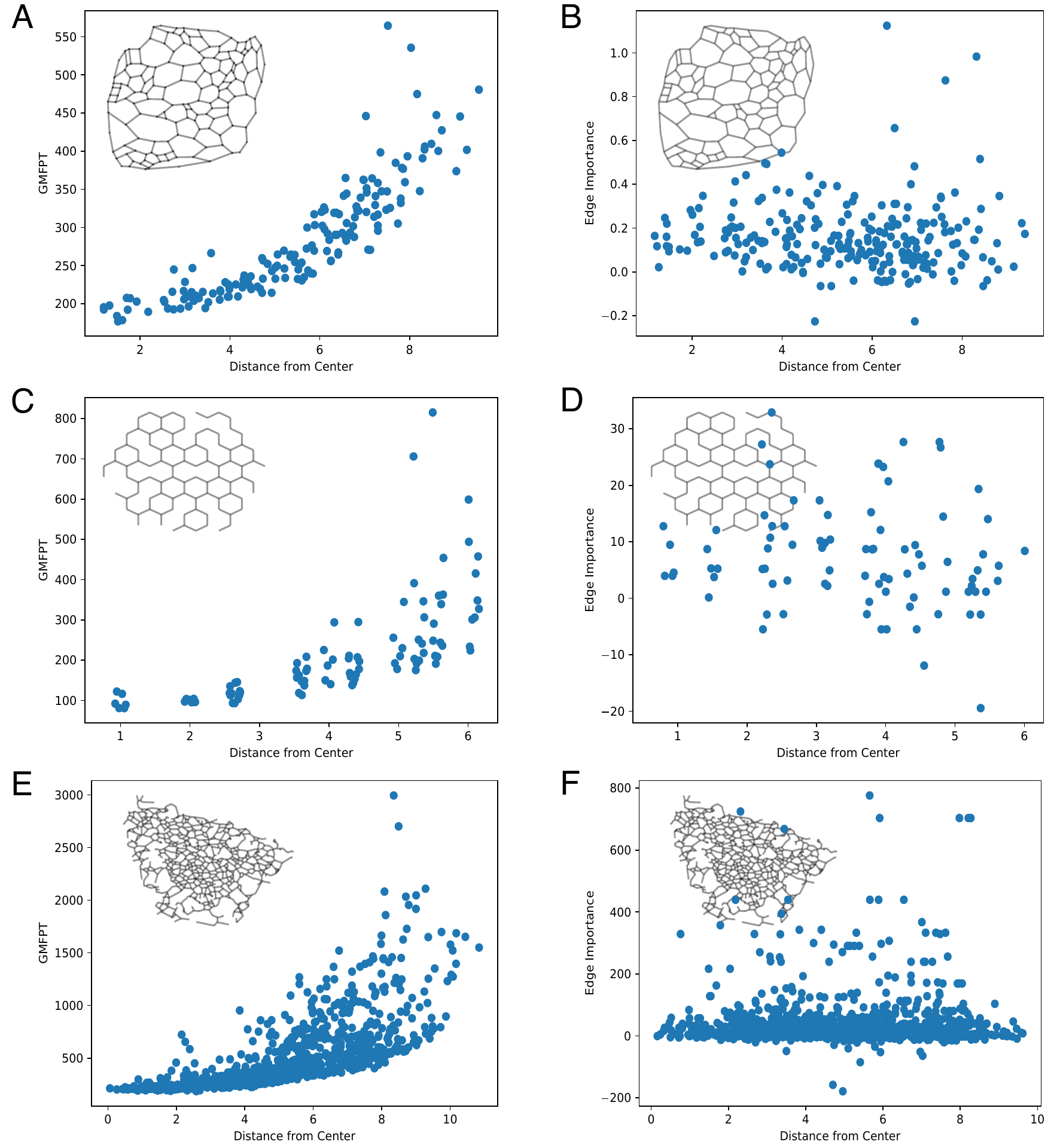}
    \caption{\edittwo{Centrality analysis of individual node or edge. (A, C, and E) Centrality of individual node compared to its GMFPT $C_r$ for the Voronoi network with 188 nodes (A), the augmented honeycomb network with 92 nodes (C), and an S2 cell (E). (B, D, and F) Centrality of individual edge compared to its relative importance for the Voronoi network with 188 nodes (B), the augmented honeycomb network with 92 nodes (D), and an S2 cell (F).}}
  \label{fig:FIG06}
\end{figure}
We used two separate sets of data obtained from experiments on \emph{D.~melanogaster}. First, we examined a series of experimental ER images from an S2 cell obtained as described in \cref{app:S2}. We demonstrate an example of the ER network extracted from a snapshot (\cref{fig:FIG03}A). \Cref{fig:FIG03}B shows that the S2 cell has a complicated ER network consisting of different types of regions. The region around the center of the ER network is dense with nodes highly connected to each other whereas the region close to the boundary of the network is relatively sparse. As a result, nodes in the inner region have a smaller impact on the MFPT compared to those in the outer region (\cref{fig:FIG03}C). We observe a similar trend in changes in edges in the inner region. That is, removing edges in the highly connected region would have a minimal effect on the calculated MFPT. However, in contrast to the nodes, edges with the highest importance correspond to those that bridge the gap between regions (see the left bottom corner in \cref{fig:FIG03}D). 
\begin{figure}[!htbp]
    \centering
    \includegraphics[width=\textwidth]{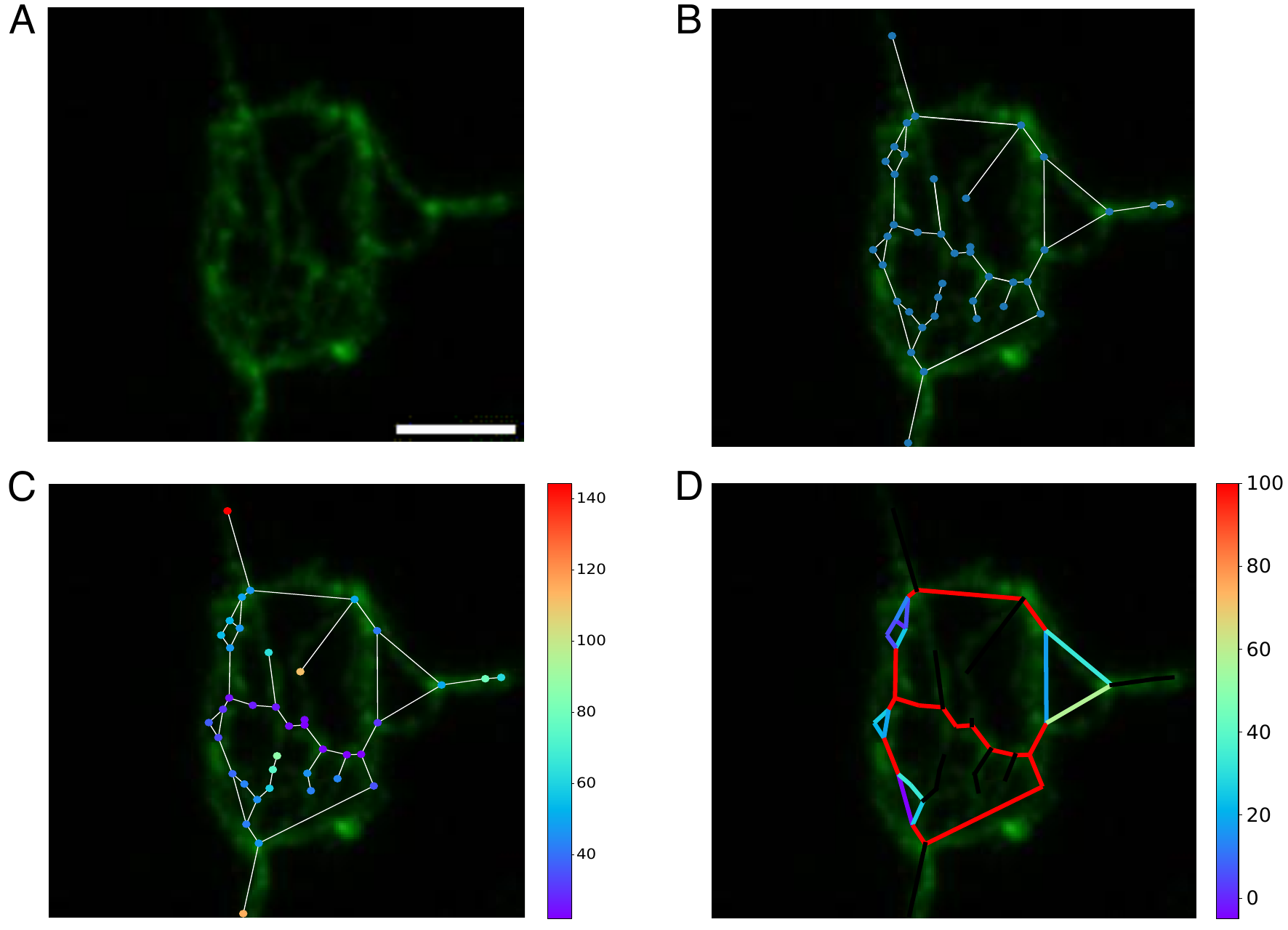}
    \caption{Diffusive transport analysis of the NMJ. (A) Original microscopy image. (B) The corresponding extracted network. (C) \edit{GMFPT}. (D) Edge importance. \edit{A 2 micron scale bar is plotted at the right bottom corner in panel (A). In panel (D) the critical edges are plotted in black (see also \cref{app:S3}).}}
  \label{fig:FIG04}
\end{figure}

To further quantify the structural properties of the S2 cell ER network, we performed a centrality analysis, which relates the distance of a node or an edge from the center of the network to its impact on the MFPT. In comparing centrality of a node in the S2 cell ER network to the synthetic networks, we found that the S2 cell ER network (\cref{fig:FIG06}E) is most similar to the Voronoi network (\cref{fig:FIG06}A), which is designed to maximise the degree of each node. It is also similar due to the large clusters of porous areas across each network. For all networks there is a positive relationship between $C_r$, \edit{GMFPT}, and the distance from the center of the network. In examining the structure of each network, we see that there are fewer edges connected to nodes on the periphery of the network (i.e. located far from the center of the network) than nodes close to the center of the network. As a result, it takes a longer time to reach nodes closer to the boundary.
\begin{figure}[!htbp]
    \centering
    \includegraphics[width=0.7\textwidth]{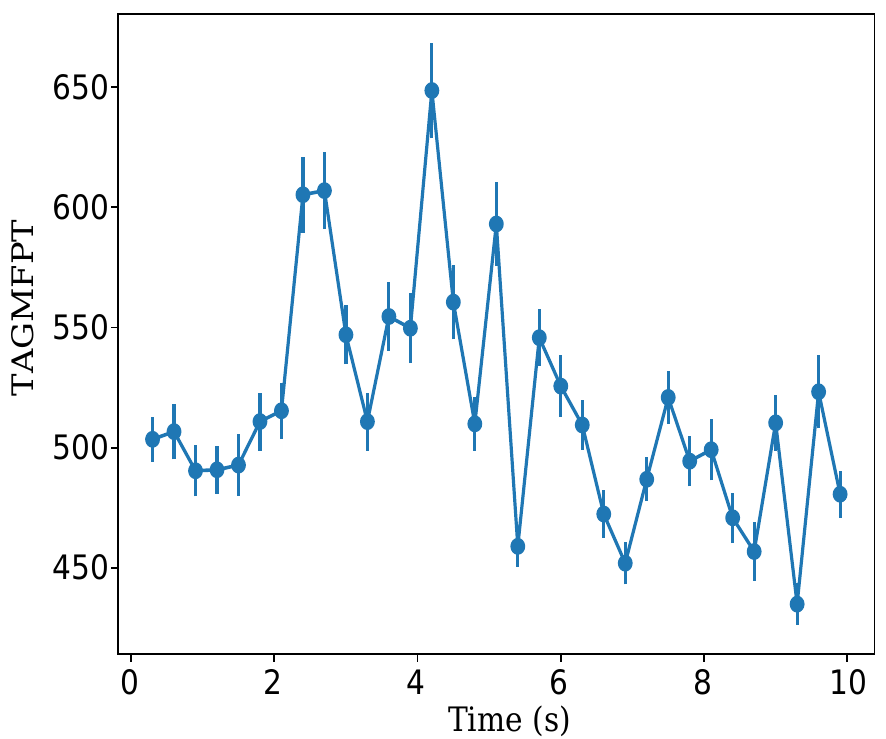}
    \caption{\edit{The TAGMFPT of the S2 cell as a function of time, with 95$\%$ confidence intervals for each data point  calculated by averaging over total number of nodes.}}
  \label{fig:FIG08}
\end{figure}

Repeating the centrality analysis for individual edges for all three types of networks (\cref{fig:FIG06}B, D, and F), we find that the S2 cell ER network demonstrates a unique relationship between edge importance and distance from center (\cref{fig:FIG06}F). In contrast to the Voronoi and the augmented honeycomb networks, we see that the edge importance for the S2 cell ER network is more heterogeneous, in that edges closer to the center have a lower importance on average. We observe that the region around the center/nucleus of the ER network is a densely connected area with short edges; therefore, we hypothesize that removing one edge would not have a significant effect in the transport process of a particle. As the distance from the center of the network is increased, we start to see an increasing number of edges that have high importance. We note that most of these edges serve as bridges that connect the dense central region of the network to its sparse boundary region. Removing such an edge could completely separate the ER network into two disconnected pieces.

The second set of ER network data we examined is taken from a \textit{D.~melanogaster} neuron at the neuromuscular junction (NMJ), with the experimental protocol described in \cref{app:NMJ}.
We apply the same routine to the NMJ ER network. We use the NMJ image of \cref{fig:FIG04}A and show the corresponding extracted network in \cref{fig:FIG04}B. The overall size of the NMJ ER network is significantly smaller and less dense than the S2 networks. Analyzing the GMFPT we find a similar trend in relation to the distance to the center as for the S2 cell ER network. We see from \cref{fig:FIG04}C that nodes that are close to the center have smaller \edit{GMFPTs} compared to those close to the boundary.

To investigate the effect of changes over time on diffusive transport in the ER network, we extracted the ER network from timelapse data from both the S2 cell and NMJ. We repeated the calculations of \cref{sect:artificialNetwork} frame-by-frame to compute the changes over time in TAGMFPT for the S2 cell (\cref{fig:FIG08}).
\begin{figure}[!htbp]
    \centering
    \includegraphics[width=\textwidth]{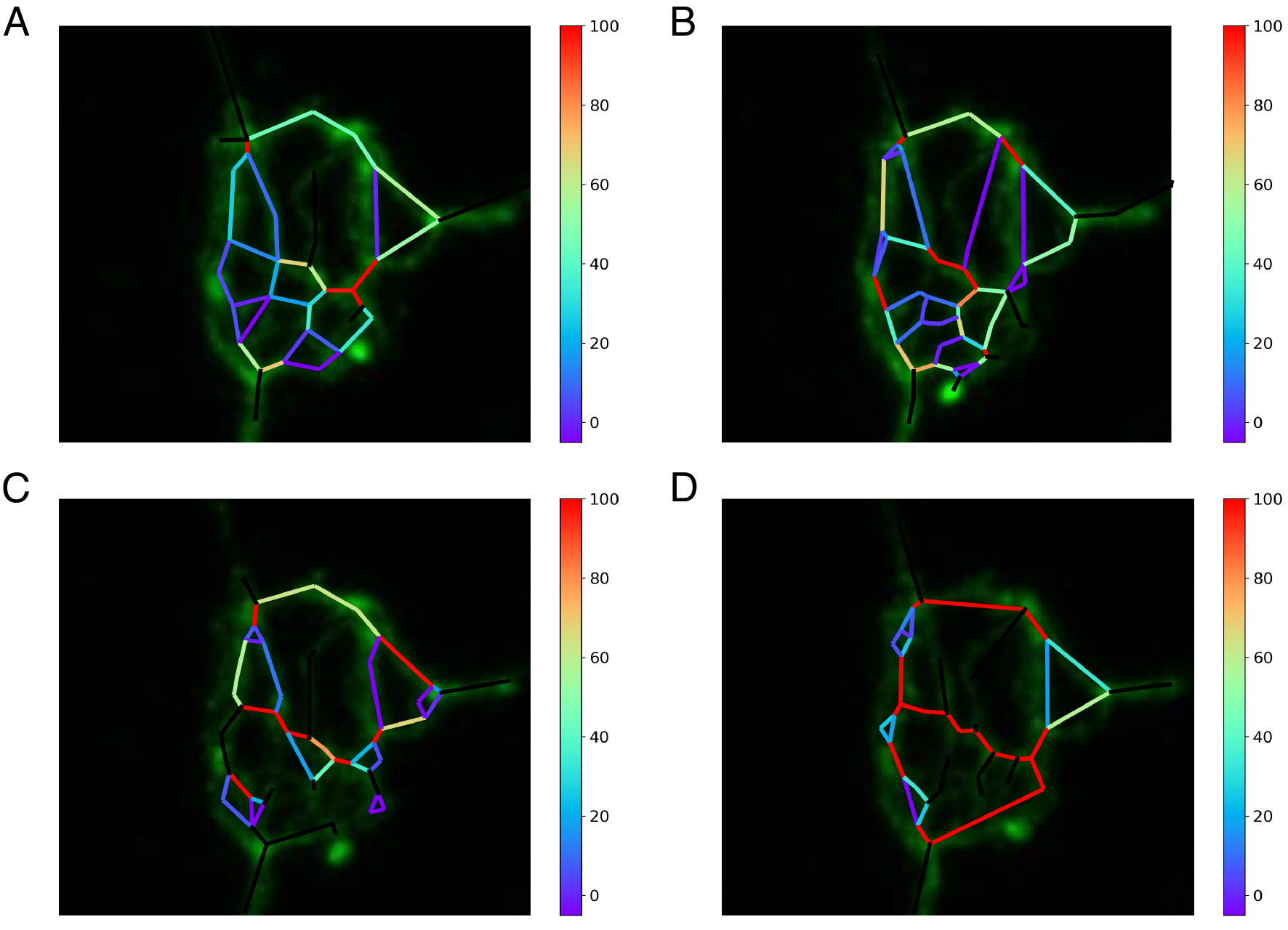}
    \caption{Edge importance over time taken from a timeseries of the NMJ. (A) 1$^\text{st}$ frame, (B) 10$^\text{th}$ frame, (C) 50$^\text{th}$ frame, and (D) 100$^\text{th}$ frame. \edit{In all panels the critical edges are shown in black.}}
  \label{fig:FIG05}
\end{figure}
In addition, we analyze the changes in the edge importance over time computed for the NMJ by comparing the $1^\text{st}$, $10^\text{th}$, $50^\text{th}$, and $100^\text{th}$ frames (\cref{fig:FIG05}). We see that most edges, except the boundary edges, have a low importance value initially (\cref{fig:FIG05}A and B). In this particular example, as time increases the ER network becomes more fragmented and edge importance increases as indicated by the additional black edges that appear in \cref{fig:FIG05}C and D. This is due to the fact that removing any of those edges  would disconnect one or more nodes from the rest of the network.
\section{Discussion}
\label{sect:conclusions}
In this work we have developed a coarse-grained model via random walks along with an optimized method to simulate the diffusive transport on networks that are evolving over time. We describe how the mean first passage time may be used to quantify the influence of the changing network morphologies on diffusive transport. Whereas on a static network our approach in calculating the MFPT is similar to several past studies of transport on spatial networks \cite{Brownetal2020,Zubenelgenubietal2021}, we describe an efficient computational method that allows us to study how the diffusive transport is affected by the evolving network structure.

After benchmarking our method using various synthetic networks, we apply it to ER network data from S2 cells and the NMJ in \emph{D.~melanogaster}. We find that the GMFPT appears to demonstrate a similar trend in both cases: 1) nodes that are closer to the center tend to be more connected and therefore have a lower GMFPT 2) nodes that are closer to the boundary tend to have a higher GMFPT as they are much harder to reach by diffusion. The edge importance reveals more information about the effect of the overall network properties of the ER on diffusion. The S2 cell network has two regions with different node densities that are bridged by relatively few edges. As a result, these edges have a higher edge importance as deleting one will separate the whole network into disconnected components with some inaccessible to the particle.

In reviewing the time evolution of the centrality of the NMJ networks, we report an interesting relation between edges with negative edge importance and their locations. \Cref{fig:FIG05} shows that these edges tend to form small loops and are connected to critical edges with high edge importance. \edit{The observation of negative edge importance is an instance of Braess' paradox \cite{braess1968paradoxon}, the counterintuitive finding that decreasing network capacity can under certain conditions lead to faster average transit times. Braess' paradox has been observed to occur in road networks \cite{youn2008price} as well as physics systems such as electronic circuits \cite{cohen1991paradoxical}, and given that similar modeling assumptions (e.g. Kirchoff's laws) are used in the present study, it is not altogether surprising that it may occur in the present context as well. However, studying the prevalence with which Braess' paradox occurs in biological networks and analyzing its consequences for transport efficiency would be an interesting avenue for further study.}

\begin{figure}[!htbp]
    \centering
    \includegraphics[width=\textwidth]{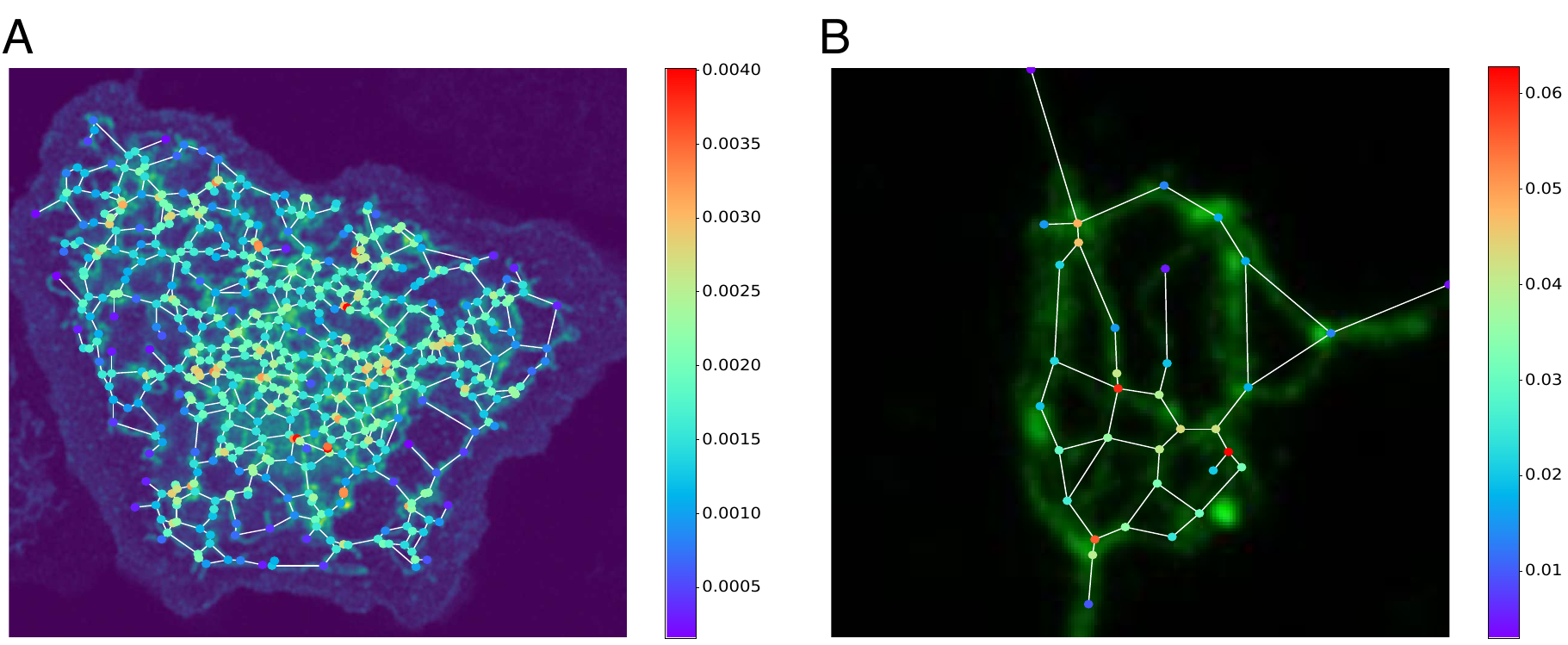}
    \caption{\edit{Stationary distribution of the MFPT for each node in ER networks. (A) 1$^\text{st}$ frame of the S2 cell timeseries. (B) 1$^\text{st}$ frame of the NMJ timeseries.}}
  \label{fig:FIG10}
\end{figure}

\begin{figure}[!htbp]
    \centering
    \includegraphics[width=\textwidth]{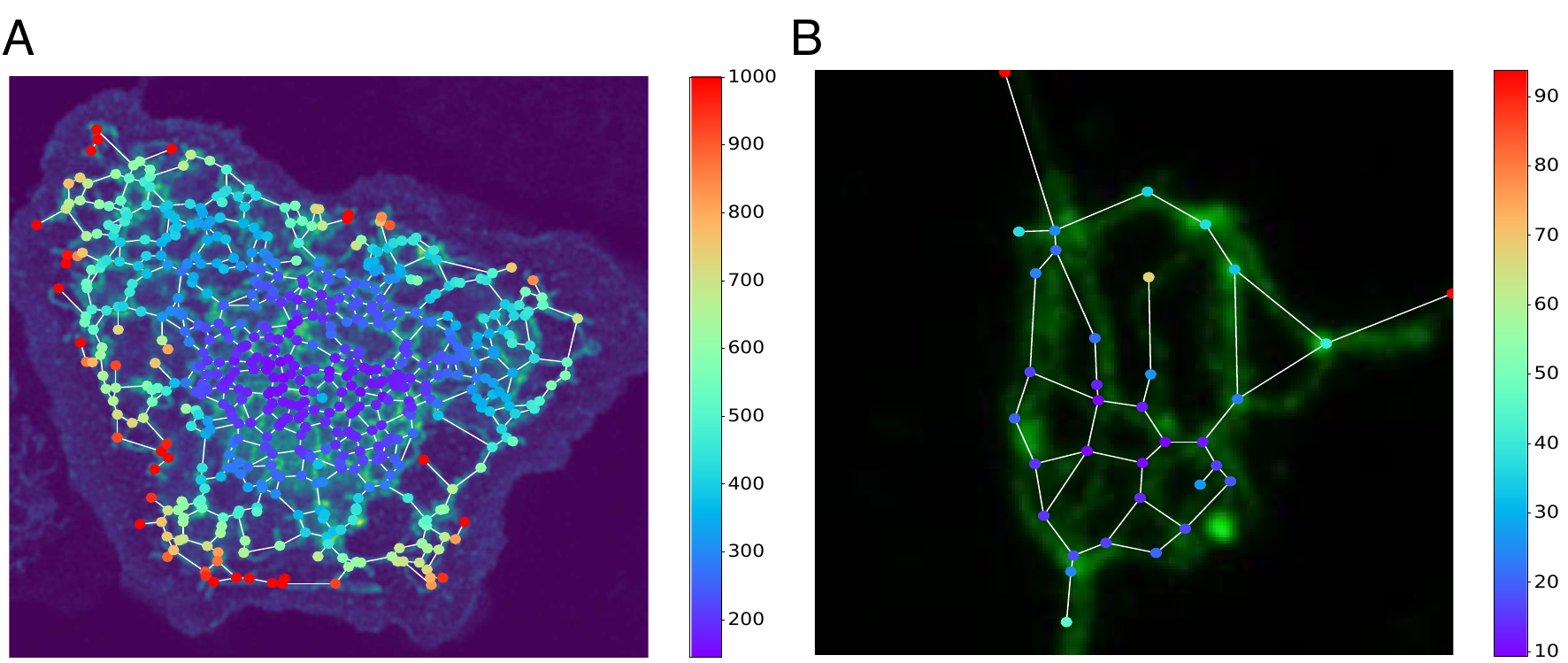}
    \caption{\edit{GMFPT of ER networks obtained using a nonuniform stationary distribution weighting. (A) 1$^\text{st}$ frame of the S2 cell timeseries. (B) 1$^\text{st}$ frame of the NMJ timeseries. Each node is weighted by its corresponding stationary distribution shown in \cref{fig:FIG10}.}}
  \label{fig:FIG09}
\end{figure}

\begin{figure}[!htbp]
    \centering
    \includegraphics[width=\textwidth]{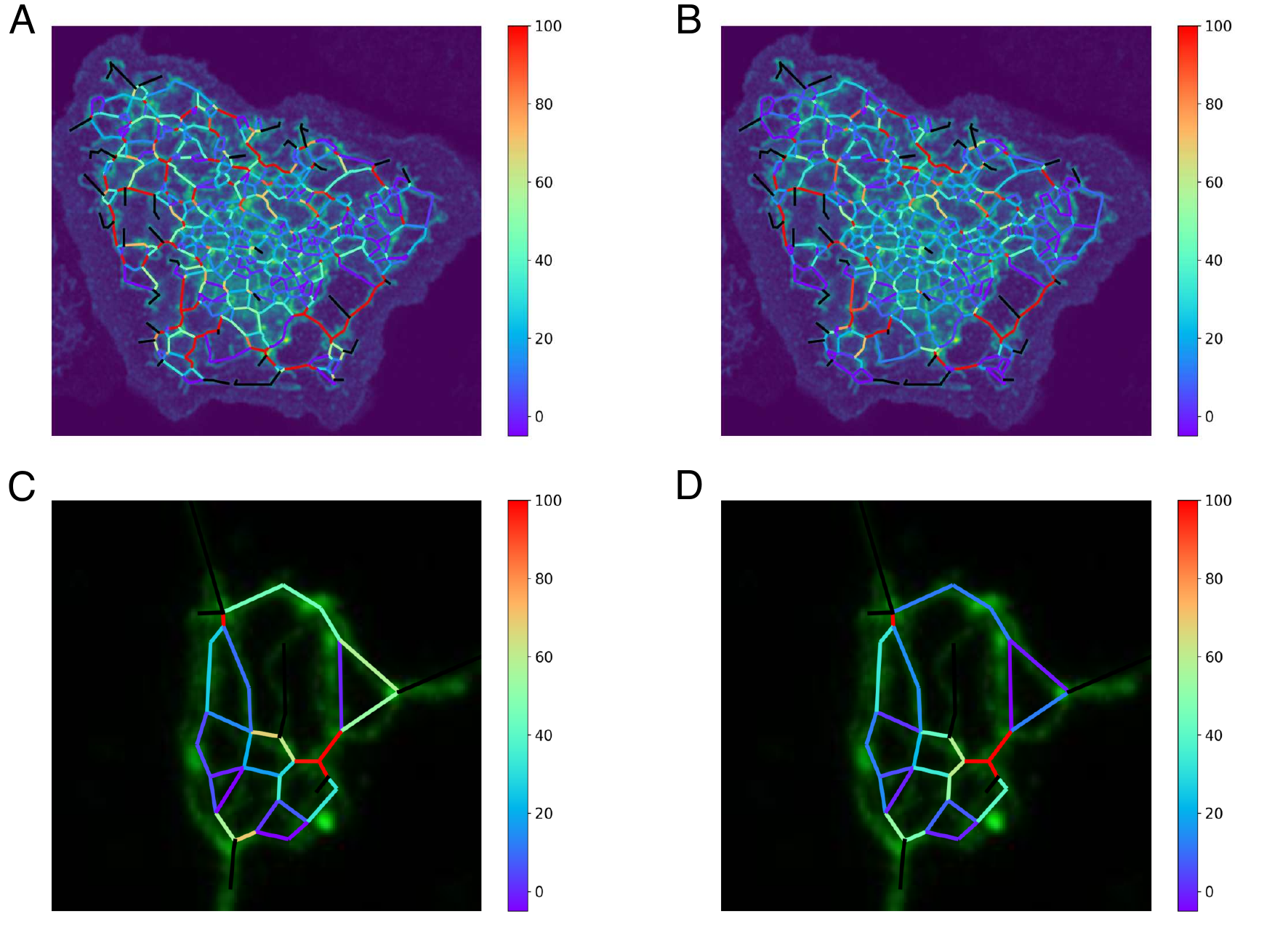}
    \caption{\edit{Comparison of edge importance of ER networks using different weightings. (A) 1$^\text{st}$ frame of the S2 cell timeseries using the uniform weighting. (B) 1$^\text{st}$ frame the S2 cell timeseries using the nonuniform stationary weighting. (C) 1$^\text{st}$ frame of the NMJ timeseries using the uniform weighting. (D) 1$^\text{st}$ frame of the NMJ timeseries using the nonuniform stationary weighting. The critical edges are plotted in black.}}
  \label{fig:FIG11}
\end{figure}

Whereas the GMFPT's discussed above are obtained by averaging with equal weight over the MFPT, one could alternatively average with respect to the stationary distribution (\cref{fig:FIG10}). This yields the expected MFPT for initial conditions chosen according to the stationary distribution, i.e.~the steady-state distribution for particles diffusing over the network (\cref{fig:FIG09}). As shown in \cref{fig:FIG11}, the qualitative features of the edge importance are not significantly changed by this choice of weighting.

\edit{We remark that although here we apply the edge importance to biological networks that change over time, the method is not inherently tied to dynamics and may be applied to other contexts as well, e.g.~to identify critical edges within transport networks. This is because the principle of using the Sherman-Morrison formula to efficiently compute the edge importance by removing one edge at a time applies more generally and is not limited to time-dependent networks. In addition, because the edge importance is computed independently for each edge, the algorithm may be  parallelized in a relatively straightforward manner to further accelerate the computation.}

There are a number of different directions that this research could take, including introducing drift to the transport process. Recent experiments have revealed that transport of particles on ER networks is affected by microtubule dynamics and is not purely diffusive \cite{StadlerEtAl2018}. Generalizing the methods presented here to non-diffusive transport would be a promising avenue for future research. 

\appendix
\section{Sherman-Morrison formula}
\label{app:sm}
The SM formula may be used to compute the fundamental matrices for different target conditions. That is to say we can go from $\edit{N_r}$ to $N_i \; \; \forall i \in \{1,2,3, ..., n\}$.
If we choose $\edit{r} = n$ we can use $\edit{N_r}$ as our base matrix upon which we can switch to any new target through a series of the SM formula applications. This process is slightly more complicated than the method for computing $\edit{N^\star_r}$ because $\edit{P_r} \in \mathds{R}^{(n-1)\times(n-1)}$ requires an extra row and column in order to swap in the probabilities of the new target condition. We define $\edit{\widetilde{P_r}} \in \mathds{R}^{n\times n}$ to be identical to the original $P$ except the $\edit{r}^{th}$ row and column will be all zeros, but $(\edit{\widetilde{P_r})_{r,r}} = 1$. For $\edit{r} = n$ we have: 
\begin{equation}
\widetilde{P_n} = 
\begin{pmatrix}
  \begin{matrix}
  P_n
  \end{matrix}
  & \rvline & \mathbf{0} \\
\hline
  \mathbf{0} & \rvline &
  \begin{matrix}
  1
  \end{matrix}
\end{pmatrix}.
\label{eq:P_n_tilda}
\end{equation}
For $r \neq n$, the rows and columns removed will split the $\edit{P_r}$ matrix into four block matrices along the $r^{th}$ row and column as follows:
\begin{equation}
\widetilde{P_r} = 
\begin{pmatrix}
  P^{UL}_r
  & \rvline & \mathbf{0}  & \rvline & P^{UR}_r\\
\hline
  \mathbf{0} &  \rvline & 1 &  \rvline & \mathbf{0}\\
\hline
  P^{LL}_r  & \rvline & \mathbf{0}  & \rvline & P^{LR}_r\\
\end{pmatrix},
\label{eq:P_r_tilde}
\end{equation}
where $P^{UL}_r,P^{LR}_r \in \mathbb{R}^{(r-1)\times(r-1)}$ and $P^{UR}_r,P^{LL}_r \in \mathbb{R}^{(n-r)\times(n-r)}$. We let $P_r \in \mathbb{R}^{(n-1)\times(n-1)}$ denote the original probability matrix without the r$^\text{th}$ row and column and is defined as
\begin{equation}
P_r = 
\begin{pmatrix}
  \begin{matrix}
  P^{UL}_r
  \end{matrix}
  & \rvline & P^{UR}_r \\
\hline
  P^{LL}_r & \rvline &
  \begin{matrix}
  P^{LR}_r
  \end{matrix}
\end{pmatrix}.
\label{eq:P_r}
\end{equation}

Having defined the $\edit{\widetilde{P_r}}$ matrices, we may form rank-1 updates of $\widetilde{P_n}$ to obtain $\widetilde{P_r} \; \; \forall r \in \{1,2,3..., n\}$. We define vectors $u_i$ and $v_i$ to be the $i^\text{th}$ row and column of our original $P$ except for the $i^\text{th}$ element: $(u_i)_i, (v_i)_i = \frac{1}{2}(1 - P_{i,i})$, this is to account for the 1 in the zero rows. The updates go as follows: 
\begin{equation}
    \begin{aligned}
        \widetilde{P_r} = \widetilde{P_n}- e_ru^T_r - v_re^T_r + e_nu^T_n + v_ne^T_n
    \end{aligned}
    \label{eq:P_r_derived}
\end{equation}
From here we may define $\widetilde{N_r}$, which is the original fundamental matrix except split across the same row and vector as $\widetilde{P_r}$:
\begin{equation}
\widetilde{N_r} = 
\begin{pmatrix}
  N^{UL}_r
  & \rvline & \mathbf{0}  & \rvline & N^{UR}_r\\
\hline
  \mathbf{0} &  \rvline & 1 &  \rvline & \mathbf{0}\\
\hline
  N^{LL}_r  & \rvline & \mathbf{0} & \rvline & N^{LR}_r\\
\end{pmatrix}
 = 
\begin{pmatrix}
  (I-P_r)^{-1}_{UL}
  & \rvline & \mathbf{0}  & \rvline &(I-P_r)^{-1}_{UR}\\
\hline
  \mathbf{0} &  \rvline & 1 &  \rvline &\mathbf{0}\\
\hline
  (I-P_r)^{-1}_{LL}  & \rvline & \mathbf{0}  & \rvline & (I-P_r)^{-1}_{LR}\\
\end{pmatrix},
\label{eq:N_r_tilde}
\end{equation}
and from this matrix one may derive $N_r$ through the same process used to obtain $P_r$ from $\widetilde{P_r}$.

We may now compute $\widetilde{N_r}$ as a series of rank 1 updates on $\widetilde{P_n}$ given by the vectors provided above:
\begin{equation}
    \widetilde{N_r} = (I - (\widetilde{P_n}- e_ru^T_r - v_re^T_r + e_nu^T_n + v_ne^T_n))^{-1}.
    \label{N_r_tilde_derived}
\end{equation}
Using $\widetilde{N_n} = (I - \widetilde{P_n})^{-1}$ as a base inversion, we may now apply the 4 rank 1 updates through the SM formula to compute $\widetilde{N_r}$, and by removing the $r^{th}$ row and column we get the matrix $N_r$ from which we can derive the $r^{th}$ row of our matrix $M$. Iterating over all $r\leq n$ we can derive the entire MFPT matrix. 

It is important to note that the order of the rank 1 updates matters. If one first adds $e_nu^T_n + v_ne^T_n$ before subtracting $e_ru^T_r + v_re^T_r$, it will create a non-invertible matrix halfway through the process and not be able to generate $\widetilde{N}_r$. The intuition behind this is that you are removing the old target condition before adding the new one, and thus with no target condition you cannot have mean first passage times, resulting in a non-invertible matrix.

\edit{
\section{Experimental materials and methods}
\subsection{S2 Cell Imaging}
\label{app:S2}
To generate pQUAST-BiP sfGFP HDEL, the sequence for BiP sfGFP HDEL was obtained by Dr. James McNew at Rice University and was cloned into pQUAST-attB (Addgene$\#$ 104880) by Vector Builder (Chicago, IL).

S2 cells were maintained at $28$ degree Celsius in Schneider's Drosophila Medium (Thermo Fisher) with 10$\%$ heat inactivated fetal bovine serum and $50$ units of penicillin-streptomycin per mL. Cells were transfected using Effectene (Qiagen, Hilden, Germany) following the manufacturer’s instructions, with pAC-7-QFBDAD (Addgene$\#$ 46096) to drive expression of QUAS-BiP sfGFP HDEL for 24 hours. Cells were transferred to cover glass chambers (Thermo Scientific, Waltham MA), coated with $0.5$ mg/ml Concanavalin A and left to settle for 1 hour at room temperature. Cells were then imaged at room temperature with a 63X (NA1.4) oil immersion objective, using an Airyscan LSM 880 microscope in super-resolution mode and Zen Black software. To record ER dynamics of \emph{Drosophila} S2 cells, 100 frames of a single slice were obtained at a frame rate of 0.65 seconds. For analysis, images were corrected for photobleaching using the histogram matching method with the imaging software FIJI \cite{schindelin2012fiji}. 

\subsection{\emph{Drosophila} larval neuromuscular junction}
\label{app:NMJ}
\emph{Drosophila melanogaster} were cultured on standard medium at 25 degree Celsius. To label neuronal ER, Vglut-Gal4 (Bloomington Drosophila Stock Center (BDSC) $\#$24635) flies were crossed with UAS BiP sfGFP HDEL(BDSC $\#$64748) flies. 3rd instar wandering larvae were dissected in Ca 2+ -free HL3.~1 \cite{feng2004modified} and axons were severed from the central nervous system. Larvae were dissected in glass slides with Press To Seal Silicone Isolators (Grace Bio-labs; CQS-13R-2.0) filled at their centers with Krayden Dow Sylgard 184 Silicone (Thermo Fisher Scientific). Small metal pins cut with nail clippers were used to stretch the larvae and the pins were pressed into the cured Sylgard 184 Silicone. Finally, a coverslip was placed on top of the dissected larvae and excess HL3.~1 was removed before imaging. Larvae were then imaged at room temperature with a 63X (NA1.4) oil immersion objective, using an Airyscan LSM 880 microscope in super-resolution mode and Zen Black software. To record ER dynamics of \emph{Drosophila} nerve terminals, imaging was performed by taking Z-stacks of 8 slices with 0.5 micron spacing for 100 frames at a frame rate of 0.73 seconds. Next, images were corrected for photobleaching and muscle contraction with FIJI using histogram matching and the ``StackReg'' plugin \cite{thevenaz1998pyramid}, respectively.

Finally, maximum projections of images were generated for analysis. The nodal and edge information are extracted using MATLAB. To extract the network information for individual frames, we begin by denoising the individual image through gridded interpolation followed by a background subtraction and guided filtering. An anisotropic diffusion filter and a threshold adjustment are applied to the denoised image to enhance the tubular structure. To obtain the corresponding image skeleton, we make use of the MATLAB functions \texttt{bwmorph} and \texttt{bwconncomp} to extract the maximum connected component of the skeleton. The resulting binary skeleton is converted to a graph to obtain the nodal and edge information for each image.

\section{Critical edges of ER networks}
Critical edges, defined as the edges whose removal would disconnect the network into multiple components, are shown for two representative ER networks in \cref{sfig:sFIG1}.
\label{app:S3}
\begin{figure}[!htbp]
    \centering
    \includegraphics[width=\textwidth]{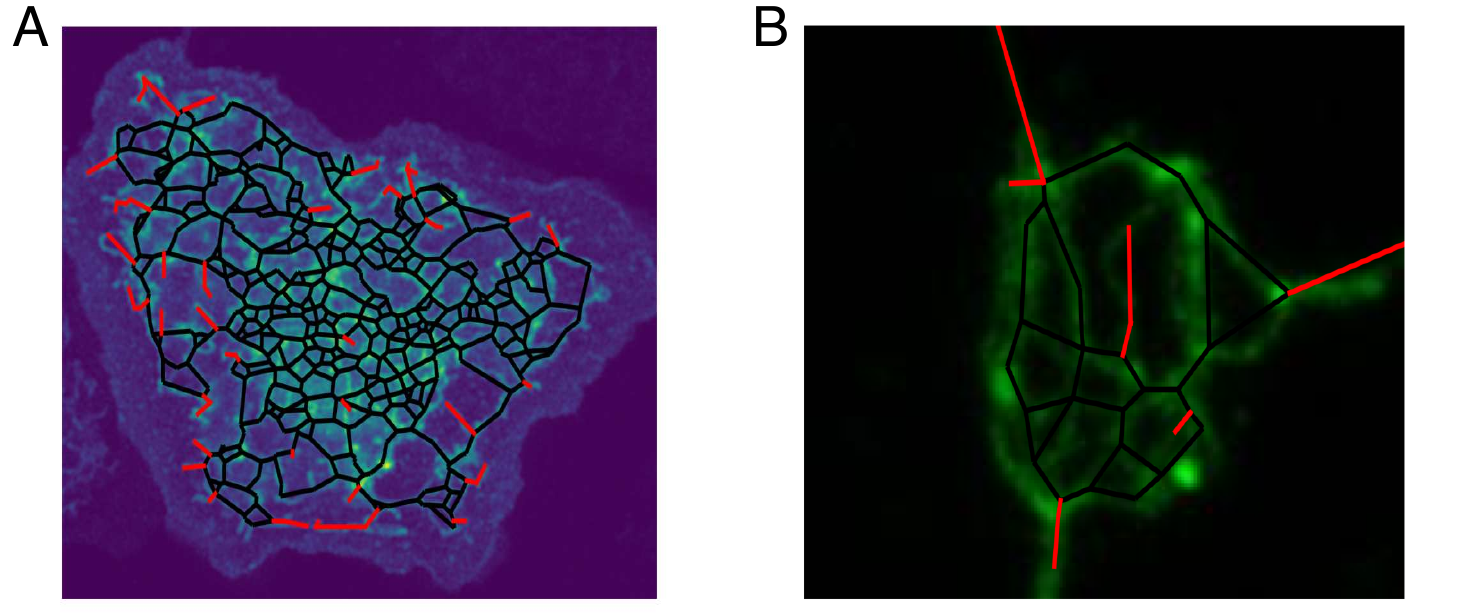}
    \caption{\edit{Critical edges of ER networks. (A) 1$^\text{st}$ frame of the S2 cell timeseries. (B) 1$^\text{st}$ frame of the NMJ timeseries. The critical edges are shown in red.}}
  \label{sfig:sFIG1}
\end{figure}
}

\section*{Acknowledgments}
We acknowledge helpful discussions with Steven Petteruti.


\end{document}